\documentclass[preprints,article,accept,moreauthors,pdftex]{Definitions/mdpi} 

\firstpage{1} 
\pubvolume{1}
\issuenum{1}
\articlenumber{0}
\pubyear{2021}
\copyrightyear{2020}
\datereceived{} 
\dateaccepted{} 
\datepublished{} 
\hreflink{https://doi.org/} % If needed use \linebreak
\pdfoutput=1

\usepackage{subfigure}
\Title{Food Odor Recognition via Multi-step Classification}

\Author{Ang Xu $^{1}$, Tianzhang Cai $^{1}$, Dinghao Shen $^{2}$, Asher Wang $^{3*}$}

\AuthorNames{Ang Xu, Dinghao Shen, Tianzhang Cai,  Yifan Hu, Yong Li, and Chang-ching Tu}

\AuthorCitation{Xu, A.; Cai, T.; Shen, D.;Hu, Y.;Li, Y.;Wang, A.;Bao, L.;Tu, C.}
\address{%
$^{1}$ \quad Shanghai Jiao Tong University; \{330861221, tztsai\}@sjtu.edu.cn

$^{2}$ \quad University of Michigan; jackyy@umich.edu

$^{3}$ \quad Bosch in China; fixed-term.Asher.WANG@cn.bosch.com}
\corres{Correspondence: fixed-term.Asher.WANG@cn.bosch.com;}

\abstract{Predicting food labels and freshness from its odor remains a decades-old task that requires a complicated algorithm combined with high sensitivity sensors. In this paper, we initiate a multi-step classifier, which firstly clusters food into four categories, then classifies the food label concerning the predicted category, and finally identifies the freshness. We use BME688 gas sensors packed with BME AI studio for data collection and feature extraction. The normalized dataset was preprocessed with PCA and LDA. We evaluated the effectiveness of algorithms such as tree methods, MLP, and CNN through assessment indexes at each stage. We also carried out an ablation experiment to show the necessity and feasibility of the multi-step classifier. The results demonstrated the robustness and adaptability of the multi-step classifier.}


\begin{document}
%%%%%%%%%%%%%%%%%%%%%%%%%%%%%%%%%%%%%%%%%%

%The order of the section titles is: Introduction, Materials and Methods, Results, Discussion, Conclusions for these journals: aerospace,algorithms,antibodies,antioxidants,atmosphere,axioms,biomedicines,carbon,crystals,designs,diagnostics,environments,fermentation,fluids,forests,fractalfract,informatics,information,inventions,jfmk,jrfm,lubricants,neonatalscreening,neuroglia,particles,pharmaceutics,polymers,processes,technologies,viruses,vision

\section{Introduction}

\subsection{Background}
    Machine olfaction is an advanced technology that captures and identifies odorous objects by distinguishing the differences according to descriptors. To extract descriptors from an object, electronic noses incorporating arrays of gas sensors and odor identification algorithms are used to imitate the process of biological noses\ \cite{webb}. 
    
    Although how to decode the psychological dimensions of human odor perception has always been a cardinal problem in olfactory research, scientists have been devoted to figuring out some comprehensive standards for the measurement and prediction of odor quality characteristics \cite{10.1093/chemse/bjs141}. Moreover, odor classification has already been incorporated in many industrial tasks. For example, meat quality can be assessed through fast detection methods \cite{meatqua}, which are essential for appropriate product management.
    
    Methods of the odor classification can mainly be grouped into two categories, which are linear and non-linear. Linear methods like principal component analysis (PCA), linear discriminant analysis (LDA), etc., were frequently used in odor identification or classification \cite{linearmethod}. However, these methods usually cannot achieve good performance because of the high non-linearity behind the model. Non-linear methods such as artificial neural networks (ANN) \cite{831508} and convolutional neural networks (CNN) \cite{s21030832} have been well-developed in recent years. Nonetheless, the applications or demonstrations of odor classification via neural network were not sufficient enough to make a holistic evaluation of the performance. Unlike linear methods that limits the number of training data \cite{lineardata}, non-linear methods usually require a large volume of a dataset for training and validation, which  enormously increases the difficulty when carrying out the experiment.
    
\subsection{Related Work}
    Odor identification and description have been highly regarded in recent years. However, most of the researches focuses on the modeling of quantitative structure upon a specific odor instead of classifying materials that compose multiple types of odorous molecules.
    
    Relevant work primarily investigated quantitative structure-odor relationship, machine olfaction, and simply odor labeling. Sensor-based machine olfaction with a neurodynamics model of the olfactory bulb uses a different color to represent the responding to different smell molecules \cite{machineOlfaction}. After signal processing, smell patterns are generated for recognition. The patterns could illustrate the concentration of the smells.
    
    Wang et al. developed a miniaturized electronic nose system to assess food freshness in refrigerators in real-time \cite{realtimeassess}. The system consists of a gas sampling module and a MOS gas sensor array. The model was built upon the sensor array results and a comparison with human sensory evaluation results.
    
    Recently, Wen et al. have proposed an Odor Labeling Convolutional Encoder-Decoder (OLCE) for Odor Sensing in Machine Olfaction \cite{olce}.  In the study, odors from seven non-crushed Chinese herbal medicines were collected by PEN-3 electronic nose. The OLCE model was built, trained, and tested by self-collecting gas response datasets. However, due to the limitation of sample types and numbers, it may not be widely applicable.

\subsection{Problem Description}
    Our work is guided by the recognition of odorous food through a classification algorithm where inputs are measured from the Bosch BME688 electronic nose. 
    
    To maximize the utility, we aimed at not only recognizing the food type but also detecting the food freshness. If completed successfully, this innovative achievement could be applicable in industrial cases such as the design of a smart fridge. Consumers would get support with the cutting-edge functionality when sifting out rotten food or evaluating the food rancidity.

    In this paper, the remaining parts are organized as follows. In section 2, we illustrated the required materials, specimens, and environment relevant to the experiment. In section 3, we introduced our multi-step classification concept and talked about how classifiers were selected in each stage. In section 4, we depicted our experimental evaluation and analyzed the results. In sections 5 and 6, we discussed the prospects of our design and formulated our conclusions.
%%%%%%%%%%%%%%%%%%%%%%%%%%%%%%%%%%%%%%%%%%
\section{Materials and Experiment Setup}

    \subsection{Flow chart}
\begin{figure}[H]
    \centering
    \subfigure{
    \includegraphics[width=6cm,height=4cm]{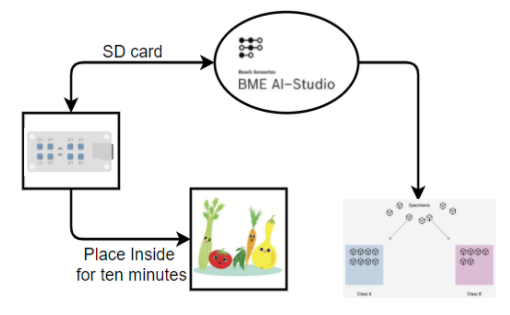} 
}
    \subfigure{
    \includegraphics[width=6cm,height=4cm]{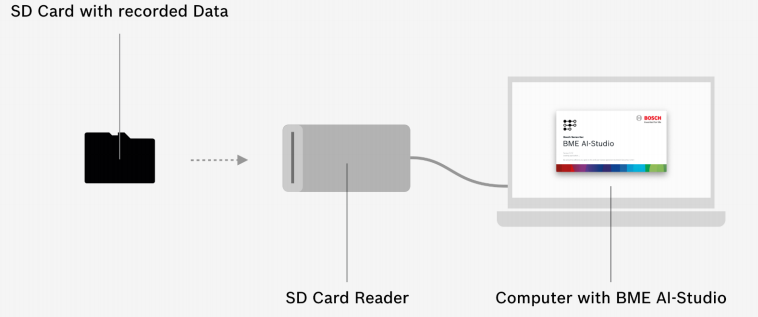}
}
    \caption{Freshness detection process for specimen}
    \label{fig:expectedProducts}
\end{figure}

        The process for detection should be a confluence of hardware for data collection and software for classification (Figure.~\ref{fig:expectedProducts}), which consists of two parts. The first part requires customers to place the gas sensor adjacent to the food or drink for detection and the SD card embedded in the sensor will extract the corresponding features as input. In the second step, recorded data are imported from the SD card to the BME studio, which will automatically execute the pre-trained model and then display the outcomes of predicted labels and freshness. This elaborately end-to-end designed system would be efficient especially for freshness tests in routine cases.

    \subsection{Materials}
        The tools used in the experiment include a mini-fridge, a processor (computer), a portable charger, a USB cable, and a development kit that incorporated electronic noses and an SD card. The computer was installed with BME studio to extract raw data. As the selected electronic nose, the BME688 gas sensor can distinguish different gas compositions by measuring unique electric fingerprints and features high sensitivity and selectivity \cite{bme}. This works on the principle that molecules going into the sensor area are charged either negatively or positively, which imposes on the electric field inside the sensor directly. The sensor provides straightforward customization for specific cases, such as detection of spoiled food and air quality \cite{bosch}. The sensor board had a heater profile, which will be run through by the sensor during a scanning cycle.  A brief setting and the configuration of the sensor board are illustrated in the figure below.

\begin{figure}[H]
    \centering
    \subfigure{
    \includegraphics[width=8cm,height=5cm]{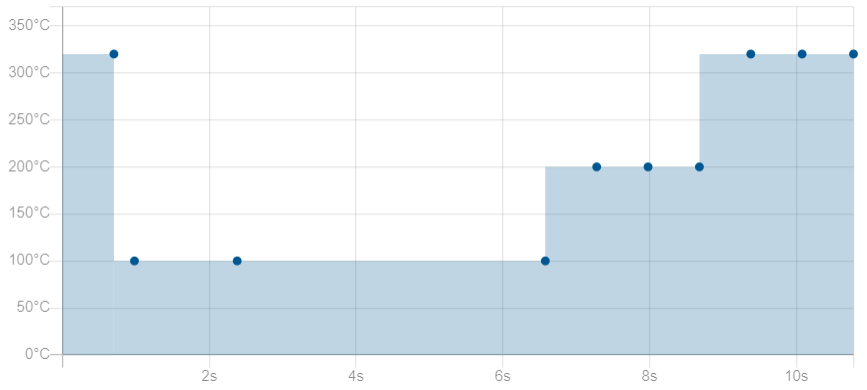} 
}
    \subfigure{
    \includegraphics[width=8cm,height=3cm]{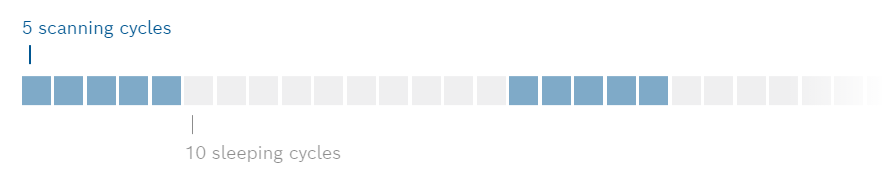}
}
    \caption{BME Board configuration (Upper: Settings of the heater profile, Down: Information of a duty cycle)}
    \label{fig:cycle}
    
\end{figure}

     The SD card inside the dev kit starts working as soon as connected to the power. The scanning will be executed corresponding to the given heater profile and duty cycles that have been set up previously. The sensor will be heated according to the given heater profile during each scanning cycle. When the sensor is in sleep mode, it will not perform the measurement. In the experiment, we used the default setting of the board configuration, which sufficed for data training. In figure \ref{fig:cycle}, the default heater profile consists of ten measurement points, which will correspondingly generate ten sample points within a scanning cycle.
        
    \subsection{Environment}
        We were devoted to simulating a scenario for practical use, so a mini-fridge was selected as our detection environment. Each time we put one piece of the specimen into the fridge and waited for minutes to ensure that the odor volatilizes from the specimen. The sensor was then connected with a portable charger and placed nearby the specimen. The time length was about 15 minutes for each detection. After that, we opened the fridge door to disperse the remaining odor. This process lasted at least five minutes to exhaust the odor in the fridge.
        \begin{figure}[H]
            \centering
            \includegraphics[scale = 0.7]{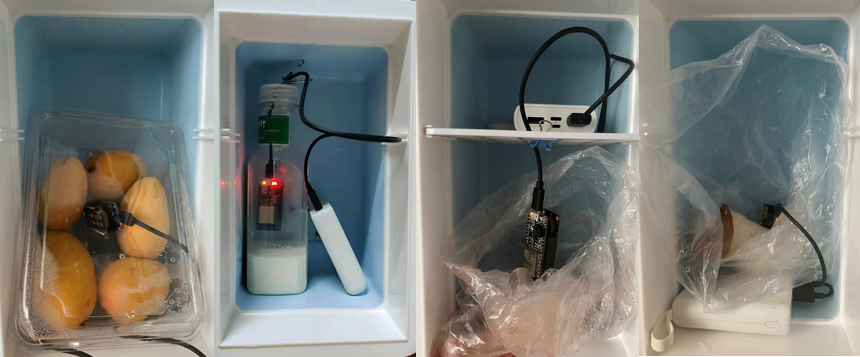}
            \caption{Environment for specimen detection (From left to right: Mango, Milk, Pork and Mushroom)}
            %\label{fig:smellPattern}
        \end{figure}
    
    \subsection{Specimens}        
         We divided our samples into four distinct categories, which were meat, vegetable, fruit, and drink. We selected food labels by popularity and  people's eating preference under each category. To enlarge the sampling flexibility, the specimens were acquired from various food markets or manufacturers. Moreover, specimens were selected in the same label with different minor types such as Fuji and Gala under the type of apple. Each specimen was repeatedly detected more than three times.
        \begin{table}[h]
            \centering
            \begin{tabular}{cccc}
            \hline Meat & Vegetable & Fruit & Drink \\
            \hline Pork & Broccoli & Apple & Coffee \\
                   Steak & Green pepper & Tangerine & Milk \\
                   Chicken meat & Mushroom & Banana & Orange juice\\
                                & Carrot & Pear \\
            \hline
            \end{tabular}
            \caption{Available labels for specimens}
            \label{tab:my_label}
        \end{table}
        
        We also separated the freshness of food into four levels for the assessment of food rancidity, which were fresh, mostly fresh, partially rotten, and rotten. For each specimen, we placed it under the normal air condition for days and recorded the status accordingly. As the decay rate was different among food and largely affected by the environment, the criterion for each level was established through life experience and the smell of food with human noses. Each of the four freshness levels is detected at least three times with different specimens under the same label. 

        %\newpage
        %\myendColumn
        \begin{figure}[H]
            \centering
            \includegraphics[scale = 0.6]{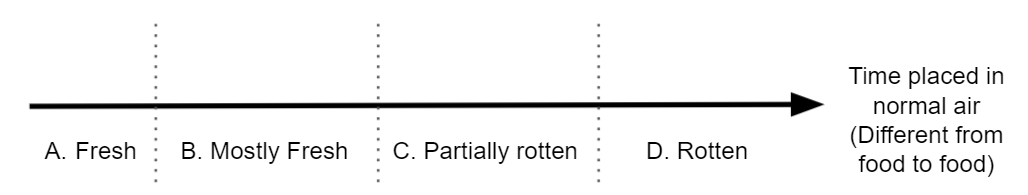}
            \caption{Separated Levels of food freshness}
            \label{fig:smellPattern}
        \end{figure}     
        \begin{figure}[H]
        	\centering
        	\subfigure{\includegraphics[scale=0.6]{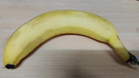}}
            \quad
        	\subfigure
        	{\includegraphics[scale=0.6]{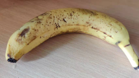}}
            \quad
        	\subfigure {\includegraphics[scale=0.6]{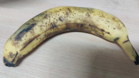}}
            \quad
        	\subfigure {\includegraphics[scale=0.6]{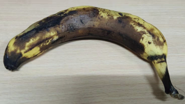}}
        	\caption{Bananas in different freshness levels (From left to right: fresh, mostly fresh, partially rotten, rotten)}
        	\label{fig:bananas}
        \end{figure}
    
%\begin{paracol}{2}
%\linenumbers
%\switchcolumn
\section{Methods}
    \subsection{Input Data Format}
        The input data from the gas sensor incorporated the detection results from bunches of food with various freshness levels. The exported data were stored in JSON format, which had 10 channels for each data point. Since dimensions like data ID, error codes were not helpful for our prediction, we extracted 4 contributory channels from one sample point, which were temperature, barometric pressure, humidity, and the inner resistance from the changing of the electric field.
        
        \begin{figure}[H]
            \centering
            \includegraphics[scale=0.7]{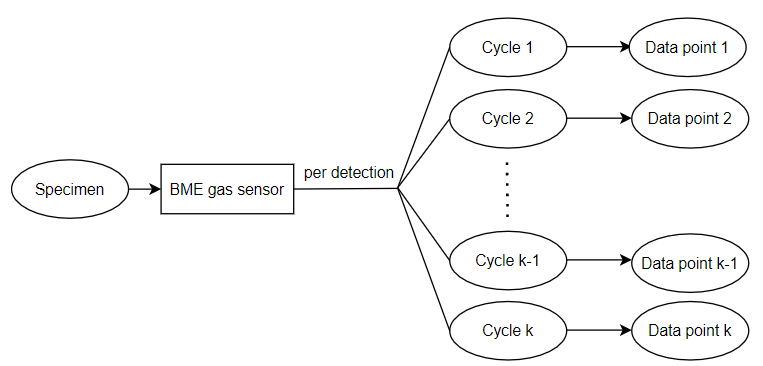}
            \caption{Generation of data points from specimens}
            \label{fig:data_point}
        \end{figure}
    
        Figure \ref{fig:data_point} describes how the input data come from the detected specimen. Here, $k$ is the number of cycles contained in a measurement session. Each cycle contains 10 data points, which was regarded as an observation, so there are $4\times10=40$ predictors for a single data point.

    \subsection{Multi-step Detection}
        %To detect the class and freshness of food, we use two-step detection method. We first use our own model to predict the food class. Then we utilize the embedded algorithm for prediction of freshness. The second step is executed through a simple neural network which is implemented after classification. We set four freshness levels, namely fresh, mostly fresh, partially rotten and rotten, which is shown in section 3.4.
        In previous studies, the freshness level was measured under the condition that the specimen's label has already been known. However, this would be inconvenient and inefficient if the food label was distinguished simply through inspection. Thereby, an end-to-end design pattern would greatly accelerate the progress for food freshness detection. Because of this, we developed a concept named multi-step classification, which consists of three distinct stages for the whole detection. \\
        
        \begin{figure}[H]
            \centering
            \includegraphics[scale=0.6]{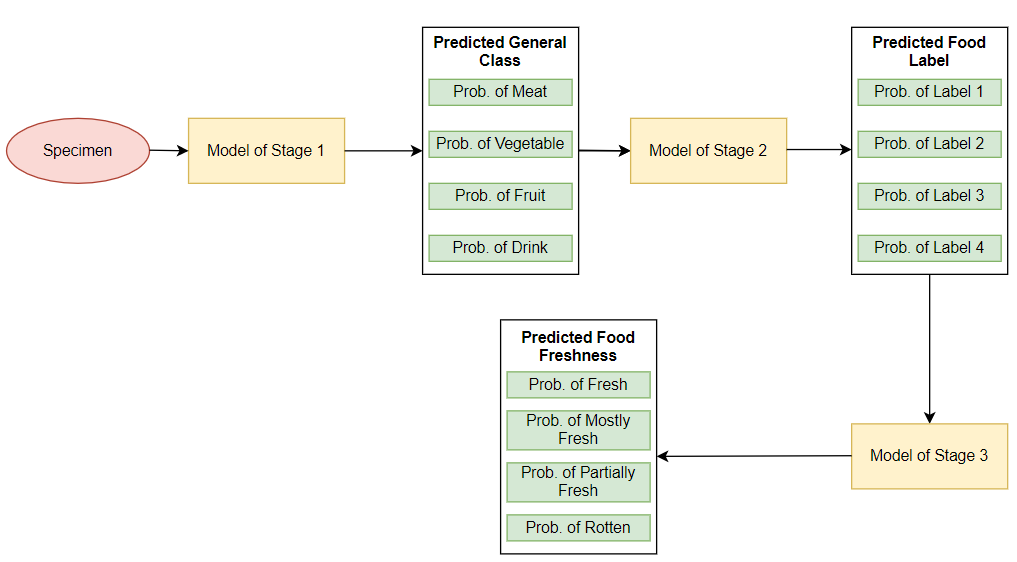}
            \caption{The design structure for the multi-step detection.}
            \label{fig:detection_process}
        \end{figure}
        
        In figure \ref{fig:detection_process}, we first took the input data of measured specimens to classify the target into four general classes, which were meat, vegetable, fruit, and drink. Due to the natural division of food categories, these general classes were supposed to be easily recognized through the dissimilarity of humidity and barometric pressure. The second stage was designated to predict the subordinate label of the specimen after the general class has been detected. For example, after correctly predicting certain specimens as fruit, we predicted whether it was an apple, banana, orange, or other kinds of fruit. In this case, we trained the model using the data underlying each general class. The implemented models embedded in the first two stages were flexible and could be self-defined. After the prediction of the label, we used a simple neural network provided by BME AI Studio to detect the freshness. In the third stage, the network was trained based on the data of a specific specimen with different freshness levels.

    \subsection{Model Selection}
        In the multi-step detection, the model implemented for each stage was under-determined. To achieve the best performance, we made attempts for a variety of classifiers at each stage. Our target was to find a combination of models in different stages to achieve the best performance. Several common algorithms that had been widely applied in the machine olfactions field were selected for comparison. 
        
        \begin{itemize}
            \item Multi-layer perception (MLP) \cite{s130302967}
            \item Decision tree (DT) \cite{AITSIALI2017145}
            \item Convolutional neural networks (CNN) \cite{SHI2019437}
            \item Support vector machine (SVM) \cite{SHI2019437}
        \end{itemize}
        
        Random Forest and AdaBoost tree classier were selected among tree methods. Random Forest constructs a multitude of decision trees through a feature bagging. An AdaBoost classifier is a meta-estimator that makes the tree tend to focus on more difficult cases by gathering information of model training difficulty and adjusting the corresponding weights of incorrectly classified instances. These methods generally outperform simple decision trees \cite{CHAN20082999}. 

        As for the prior experience, these classifiers behaved well on in-sample data. However, our model needed to be generalized to unrecognized data, which implies that the test accuracy should be emphasized. Because of the high similarity of in-sample data, only having high validation accuracy among the same specimens with different data points may bring spurious results. We required a model that could perform well on out-of-sample data as well. In view of this, we performed data preprocessing with principal component analysis (PCA) and linear discriminant analysis (LDA) \cite{pcalda}. PCA reordered the importance of predictors in sequence by projecting the input data onto a lower-dimensional space and LDA performs the dimensionality reduction by finding a linear combination of features that characterizes multiple classes. PCA can be performed before LDA to avoid over-fitting and regularize the problem in high dimensional and singular cases. \cite{YANG2003563}. 
        
        \begin{table}[h]
            \centering
            \begin{tabular}{cccc}
                \hline
                Model &  Layer &  Filter Shape & Input Size \\
                \hline
                MLP   &  Fully connected layer & 64 & 40 \\
                      &  ReLu layer            & \\
                      &  Dropout layer         & \\
                      &  Fully connected layer & 64 & 64 \\
                \hline
                CNN   & 2D convolutional layer & 2$\times$2 & 1$\times$4$\times$10  \\
                      & Max pooling layer      & 3$\times$1 & 4$\times$4$\times$10  \\
                      & 2D convolutional layer & 2$\times$2 & 4$\times$2$\times$4   \\
                      & Flatten layer          &     & 16$\times$2$\times$4  \\
                      & Dropout layer          & \\
                      & Fully connected layer  & 16 & 128 \\
                \hline
                
            \end{tabular}
            \caption{Neural network architecture for MLP and CNN models}
            \label{tab:my_label}
        \end{table}
        
\section{Results}   
    \subsection{Data Preprocessing and Visualization}
        An explicit view of the distribution of different classes and labels was given in this section. To display the data in a clear manner, we first normalized the data to be zero-centered by calculating the $i^{th}$ predictor value as $x_{i}^{\prime}$ follows:
        \begin{equation}
        x_{i}^{\prime}=\frac{x_{i}-\bar{x}}{x_{\max }-x_{\min }}
        \end{equation}
        where $x_i$ is the original input data, $\bar{x}$ is the average value of all predictors in the data point, and $x_{max}$ and $x_{min}$ are the maximum and the minimum value of the data point respectively.
        
        The normalized data contained multiple features under different timestamps. As mentioned in section 3.3, PCA and LDA were then used to visualize the data in lower dimensions. After preprocessing, we extracted 2 or 3 predictors (2 predictors for 3 classes and 3 predictors for 4 classes) with the largest importance. 
        \begin{figure}[H]
          \centering
            \subfigure{
            \label{fig:subfig:a}
            \includegraphics[width=1.5in]{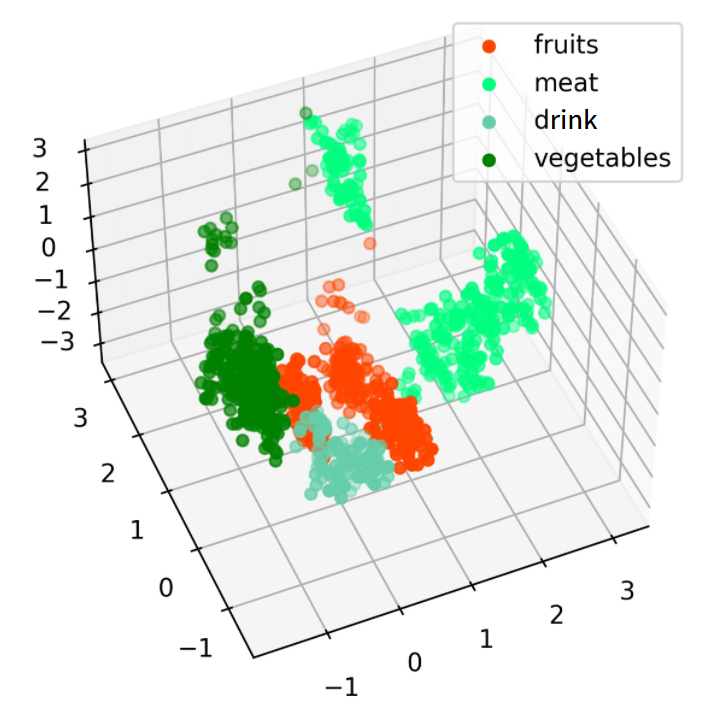}}
          \subfigure{
            \label{fig:subfig:a} %% label for first subfigure
            \includegraphics[width=1.5in]{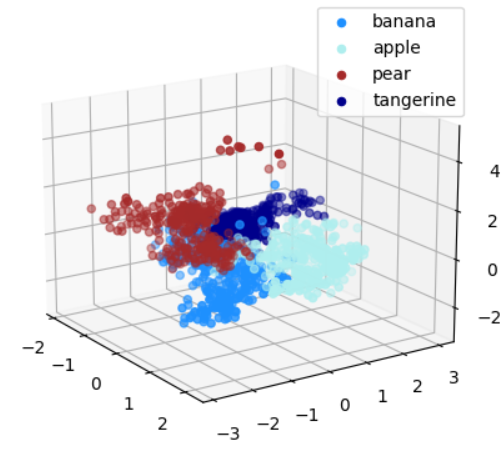}}
          \subfigure{
            \label{fig:subfig:b} %% label for second subfigure
            \includegraphics[width=1.5in]{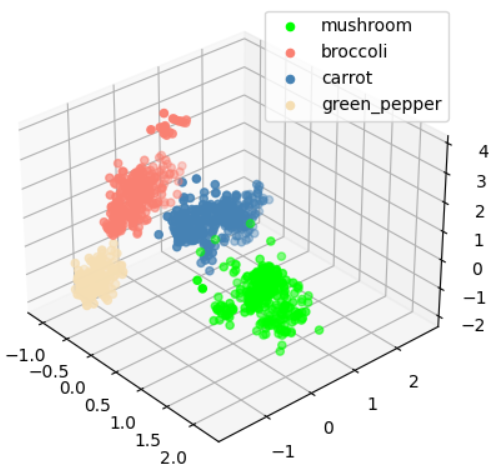}}
            
           %% label for entire figure 
          \subfigure{
            \label{fig:subfig:a} %% label for first subfigure
            \includegraphics[width=1.5in]{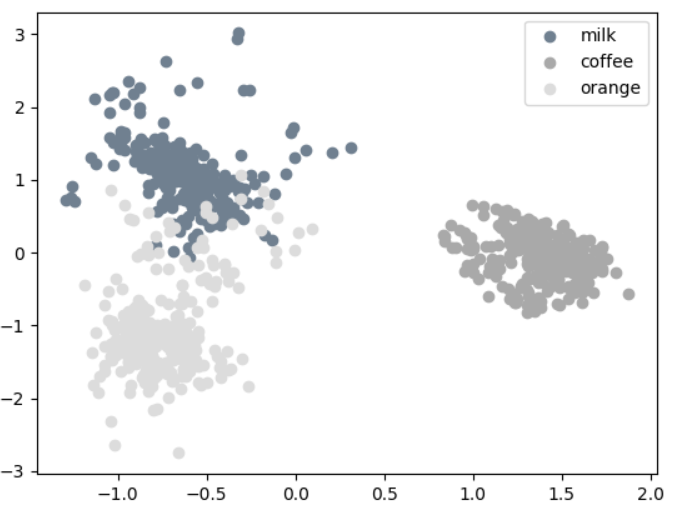}}
            \subfigure{
            \label{fig:subfig:a} %% label for first subfigure
            \includegraphics[width=1.5in]{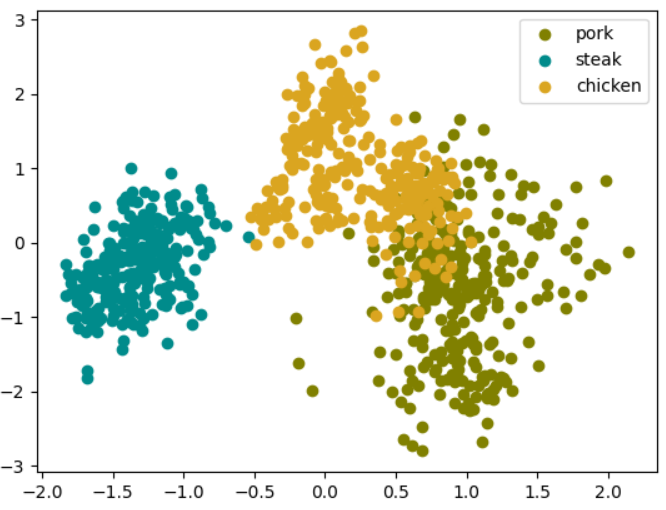}}
          \caption{Input data visualization after preprocessing (The coordinates represent different dimensions for extracted features)}
          \label{fig:pca}
        \end{figure}
        
        From figure \ref{fig:pca}, different classes were clustered together  after data preprocessing. This procedure eliminated unnecessary predictors and laid a good foundation for the classification. The visualization gave a clear view of the division of different classes in each stage. 
        
    \subsection{Evaluation of Multi-step Classification}
        For the first two-stage, we trained all selected models and chose the one with the best performance among evaluation indexes, which are accuracy, F-score, and Cohen's kappa coefficient. 
        
        The F-1 score was calculated from the mean of precision and recall rate
        \begin{equation}
        F_{1}=2 \cdot \frac{\text { precision } \cdot \text { recall }}{\text { precision }+\text { recall }}
        \end{equation}
        where the precision is the number of true positive results over the total number of positive results and the recall is the number of true positive results over the total number of samples that should have been identified as positive.
        
        We also calculated the Kappa coefficient to evaluate the inter-rater reliability \cite{kappa} for our model effectiveness. 
        \begin{equation}
        \kappa \equiv \frac{p_{o}-p_{e}}{1-p_{e}}=1-\frac{1-p_{o}}{1-p_{e}}
        \end{equation}
        
        where $p_o$ is the accuracy, and $p_e$ is the hypothetical probability of chance agreement which can be expressed as 
        \begin{equation}
        p_e = \sum_{i=1}^{k}\frac{m_in_i}{n^2}
        \end{equation}
        where $k$ is the total number of classes, $n$ is the total number of data points, $m_i$ is the number of data points that belongs to the class $i$ in the original dataset and $n_i$ is the number of data points that belongs to the class $i$ from the predicted results.
        
        Leave-one-out cross-validation was calculated to verify the effectiveness of our models. Instead of using one observation as the validation set, we left out all observations during the measurement time per detection and this avoided the similarity among data in the same measurement session. For MLP and CNN, each model was trained for 50 epochs to reach convergence.
        
        \begin{table}[H]
            \centering
            \begin{tabular}{cccc}
               \hline
               Algorithm  &  Accuracy & F-1 score & Kappa coefficient\\
               \hline
               MLP (no preprocessing) & 0.3188 & 0.2124 & 0.0361\\
               CNN (no preprocessing) & 0.3088 & 0.1707 & 0.0297\\
               MLP & 0.8635 & 0.8528 & 0.7933\\
               Random Forest & 0.8867 & 0.8897 & 0.8380\\
               SVM & 0.7894 & 0.7929 & 0.7032\\
               AdaBoostTree & 0.8397 & 0.8342 & 0.7984\\
               Logistic regression & 0.8377 & 0.8351 & 0.7769\\
               \hline
            \end{tabular}
            \caption{Performance evaluation for stage1 (classification of general classes)}
            \label{tab:general}
        \end{table}
        
        From table \ref{tab:general}, Random Forest had the best performance when predicting the general class in the first stage. It achieved an average accuracy of 0.8867, F-1 score of 0.8897, and Kappa coefficient of 0.8380. Algorithms without preprocessing got a validated accuracy lower than 0.35, which revealed the worst result.
        
        In the second stage, a model was trained for each of the four general classes. Again, we calculated the performance indexes using leave-one-out cross-validation, which averaged the performance indexes of more than ten cross-validated models. 
        
        \begin{table}[H]
            \centering
            \begin{tabular}{cccc}
               \hline
               Algorithm  &  Accuracy & F-1 score & Kappa coefficient\\
               \hline
               MLP (no preprocessing) & 0.3286 & 0.2856 & 0.0328\\
               CNN (no preprocessing) & 0.3760 & 0.2600 & 0.1670\\
               MLP & 0.9142 & 0.9344 & 0.8995\\
               Random Forest & 0.8383 & 0.8136 & 0.7844\\
               SVM & 0.9635 & 0.9627 & 0.9513\\
               AdaBoostTree & 0.9106 & 0.9057 & 0.8808\\
               Logistic regression & 0.9092 & 0.8932 & 0.9029\\
               \hline
            \end{tabular}
            \caption{Performance evaluation for the vegetable category in stage 2 (classification of vegetable classes)}
            \label{tab:vegetable}
        \end{table}
        
        \begin{table}[H]
            \centering
            \begin{tabular}{cccc}
               \hline
               Algorithm  &  Accuracy & F-1 score & Kappa coefficient\\
               \hline
               MLP (no preprocessing) & 0.3468 & 0.3008 & 0.0369\\
               CNN (no preprocessing) & 0.4791& 0.3295 & 0.2938\\
               MLP & 0.9015 & 0.8843 & 0.8744\\
               Random Forest & 0.8342 & 0.7952 & 0.7802\\
               SVM & 0.8651 & 0.8440 & 0.8194\\
               AdaBoostTree & 0.8016 & 0.7557 & 0.7354\\
               Logistic regression & 0.8879 & 0.8572 & 0.8504\\
               \hline
            \end{tabular}
            \caption{Performance evaluation for fruit category in stage 2 (classification of fruit classes)}
            \label{tab:vegetable}
        \end{table}
        
        \begin{table}[H]
            \centering
            \begin{tabular}{cccc}
               \hline
               Algorithm  &  Accuracy & F-1 score & Kappa coefficient\\
               \hline
               MLP (no preprocessing) & 0.4159 & 0.3485 & 0.0381\\
               CNN (no preprocessing) & 0.3587 & 0.2225 & -0.0048\\
               MLP & 0.8619 & 0.8318 & 0.7048\\
               Random Forest & 0.9003 & 0.8719 & 0.8500\\
               SVM & 0.8683 & 0.8506 & 0.8024\\
               AdaBoostTree & 0.8334 & 0.7699 & 0.7381\\
               Logistic regression & 0.8825 & 0.8563 & 0.8238\\
               \hline
            \end{tabular}
            \caption{Performance evaluation for drink category in stage 2 (classification of drink classes)}
            \label{tab:vegetable}
        \end{table}
        
        \begin{table}[H]
            \centering
            \begin{tabular}{cccc}
               \hline
               Algorithm  &  Accuracy & F-1 score & Kappa coefficient\\
               \hline
               MLP (no preprocessing) & 0.3158 & 0.2966 & 0.0343\\
               CNN (no preprocessing) & 0.3393 & 0.2778 & 0.1080\\
               MLP & 0.9203 & 0.9344 & 0.9195\\
               Random Forest & 0.8715 & 0.8270 & 0.8144\\
               SVM & 0.9235 & 0.9227 & 0.9130\\
               AdaBoostTree & 0.9106 & 0.9057 & 0.8808\\
               Logistic regression & 0.9392 & 0.9232 & 0.9022\\
               \hline
            \end{tabular}
            \caption{Performance evaluation for the meat category in stage 2 (classification of meat classes)}
            \label{tab:meat}
        \end{table}
        
        From table \ref{tab:meat}, SVM, MLP, Random Forest and logistic regression outperformed other classifiers on vegetable, fruit, drink and meat classifier respectively. All of them obtained an accuarcy over 0.9 with stable F-1 scores and kappa coefficients. We can infer that tested algorithms are sensitive to food categories in prediction.

    \subsection{Freshness Detection}    
        After the label has been recognized, we assessed the food freshness by training a model only using data regarding the recognized label. Due to the sparsity of the sample, we leveraged the embedded algorithm provided by BME AI Studio \cite{bme}. The algorithm features simplicity and adaptability to address the issue of lacking in sampling. The neural net architecture for the embedded algorithm was pre-defined as two fully connected layers with 16 hidden nodes for each.
        
        \begin{table}[H]
            \centering
            \begin{tabular}{cccc}
                \hline Algorithm & Accuracy & F-1 score & Kappa coefficient \\
                \hline
                       Embedded MLP algorithm & 0.9714 & 0.9328 & 0.9399 \\
                       MLP (no preprocessing) & 0.9615 & 0.9315 & 0.9201 \\
                       CNN (no preprocessing) & 0.7984 & 0.7716 & 0.7159 \\
                       Random Forest & 0.9328 & 0.9053 & 0.8982 \\
                       SVM & 0.9510 & 0.9128 & 0.9019 \\
                       AdaBoostTree & 0.9506 & 0.9123 & 0.8995\\
                       \hline
            \end{tabular}
            \caption{Performance evaluation for stage 3 (freshness detection)}
            \label{tab:freshness}
        \end{table}
    
        Table \ref{tab:freshness} shows that BME 688 gas sensor had a better performance on freshness detection than the classification of labels. The sensor was sensitive to the pungent odor from rotten food. The average accuracy reached , .
        
    \subsubsection{Ablation Experiment}    
    In this part, we compared the multi-step detection with one-step methods that directly classified the food label and its freshness level. In multi-step classification, the best classifiers were selected from each stage to calculate the overall accuracy of the whole detection process. In one-step methods, results of the food label were derived directly through the comparison classifiers in the model selection part. Then the food freshness was detected under the same model as stage 3 in the multi-step detection.
    
        \begin{table}[H]
            \centering
            \begin{tabular}{cccc}
                \hline Algorithm & Accuracy & F-1 score & Kappa coefficient \\
                \hline
                       Multi-step classification & 0.8011 & 0.7904 & 0.7716\\
                       MLP (no preprocessing) & 0.2521 & 0.2376 & 0.2480\\
                       CNN (no preprocessing) & 0.2836 & 0.2718 & 0.2574\\
                       MLP & 0.5765 & 0.4798 & 0.5275\\
                       Random Forest & 0.4872 & 0.4290 & 0.4478\\
                       SVM & 0.4362 & 0.3812 & 0.3929 \\
                       AdaBoostTree & 0.3631 & 0.3141 & 0.2757 \\
                       Logistic Regression & 0.4515 & 0.4161 & 0.4093 \\

                \hline
            \end{tabular}
            \caption{Overall performance for multi-step detection and one-step algorithms}
            \label{tab:ablation}
        \end{table}
    
    From table \ref{tab:ablation}, the overall accuracy increases from 0.5765 to 0.8011 with our multi-step classification. The overall accuracy was over 80 percent for the whole system. The F-1 score and Kappa coefficient were approaches to 0.8, which suggests high feasibility and reliability. By contrast, the performance indexes are lower than 0.6 in one-step classifiers. The experimental results improved the indexes by over 20 percent, which demonstrated the effectiveness and robustness. 
%%%%%%%%%%%%%%%%%%%%%%%%%%%%%%%%%%%%%%%%%%
\section{Discussion}
    Odor identification through electronic noses is a process of discriminating various ingredients that gas sensors are sensitive to \cite{olce}. Since for most electronic noses, the only measured variable is the resistance under different external conditions such as temperature, pressure, and humidity, a robust algorithm to distinguish a variety of responses is essential to complement the lack of descriptors brought by the sensor. 
    
    In previous studies, it is hard to detect the food type and the freshness at the same time. We resolved this problem by developing the concept of multi-step detection, which was well-designed, trainable, and not computational. The whole dataset was dismantled into lower-level data and used for training multiple times, which maximized the sample utilization ratio. The great potentiality of this design implies the possibility of the application in industrial fields. The pre-trained model can be loaded into microprocessing units to monitor the freshness level of food preserved in the fridge blocks.
    
    This design concept utilizes the natural boundaries of food species. It should also have decent performance on building other multi-class odor classifier. For example, one may keep on classifying the minor classes of a certain kind of fruit or predicting the maturity of fruits through this multi-class classifier.

%%%%%%%%%%%%%%%%%%%%%%%%%%%%%%%%%%%%%%%%%%
\section{Conclusions}
    In this paper, We developed a method of odor classification based on food smell using Bosch's BME688 gas sensor combined with machine learning algorithms. Specimens were collected under different food markets to ensure versatility and universality. Our multi-step classification managed to classify the label and further detected the freshness of food specimens efficiently. Moreover, since the embedded model is self-determined in each stage, the concept is flexible and adjustable and has significantly fewer constraints compared to other design patterns. The results demonstrated the robustness and effectiveness and the decent performance exhibited its great potentiality on industrial applications such as embedding our design into a smart fridge to detect the rancidity of food.

   % The best model we have developed is a convolutional neural network (CNN). It has a high accuracy on classifying different kinds of food and predicting its freshness level. Results demonstrated that the two-step detection with CNN had a decent performance according to the performance evaluation indexes.

%%%%%%%%%%%%%%%%%%%%%%%%%%%%%%%%%%%%%%%%%%
\vspace{6pt} 

%%%%%%%%%%%%%%%%%%%%%%%%%%%%%%%%%%%%%%%%%%
%% optional
%\supplementary{The following are available online at \linksupplementary{s1}, Figure S1: title, Table S1: title, Video S1: title.}


%%%%%%%%%%%%%%%%%%%%%%%%%%%%%%%%%%%%%%%%%%
\authorcontributions{Conceptualization, A. Xu; software, T. Cai and D. Shen; validation, T. Cai; formal analysis, A. Xu; investigation, A. Xu; data collection, A. Xu; writing, A.Xu; supervision and consultation, A. Wang; All authors have read and agreed to the published version of the manuscript.}

\funding{This research was not funded.}

\institutionalreview{Not applicable.}

\dataavailability{Not applicable.}

\acknowledgments{We would like to acknowledge the support provided by  Changching Tu (Assistant Professor at Shanghai Jiao Tong University),  Lei Bao (Senior Expert at Bosch Research), Yong Li (Undergraduate at Shanghai Jiao Tong University) and Yifan Hu (Undergraduate at Shanghai Jiao Tong University) in the process of our experiment.}

\conflictsofinterest{The authors declare no conflict of interest.} 

%% Optional

%%%%%%%%%%%%%%%%%%%%%%%%%%%%%%%%%%%%%%%%%%
%% Only for journal Encyclopedia
%\entrylink{The Link to this entry published on the encyclopedia platform.}

%%%%%%%%%%%%%%%%%%%%%%%%%%%%%%%%%%%%%%%%%%
%% Optional
\abbreviations{The following abbreviations are used in this manuscript:\\

\noindent 
\begin{tabular}{@{}ll}
MLP & Multi-Layer Perception \\
CNN & Convolutional Neural Network \\
LDA & Linear Discriminant Analysis \\
PCA & Principal Component Analysis \\
SVM & Support Vector Machine\\
DT & Decision Tree
\end{tabular}}

%%%%%%%%%%%%%%%%%%%%%%%%%%%%%%%%%%%%%%%%%%
%% Optional
\appendixtitles{no} % Leave argument "no" if all appendix headings stay EMPTY (then no dot is printed after "Appendix A"). If the appendix sections contain a heading then change the argument to "yes".
%\appendixstart
%\appendix
%\section{}
%\subsection{}

%\begin{specialtable}[H] 
%\tablesize{\scriptsize}
%\caption{This is a table caption. Tables should be placed in the main text near to the first time they are~cited.\label{tab1}}
%\tablesize{} % You can specify the fontsize here, e.g., \tablesize{\footnotesize}. If commented out \small will be used.

%\end{specialtable}

%%%%%%%%%%%%%%%%%%%%%%%%%%%%%%%%%%%%%%%%%%
%\end{paracol}
\reftitle{References}

\bibliography{egbib}
%%%%%%%%%%%%%%%%%%%%%%%%%%%%%%%%%%%%%%%%%%
%% for journal Sci
%\reviewreports{\\
%Reviewer 1 comments and authors’ response\\
%Reviewer 2 comments and authors’ response\\
%Reviewer 3 comments and authors’ response
%}
%%%%%%%%%%%%%%%%%%%%%%%%%%%%%%%%%%%%%%%%%%
\end{document}